# Tunable Nanoislands Decorated Tapered Optical Fibers Reveal Concurrent Contributions in Through-Fiber SERS Detection

Di Zheng[1,*], Muhammad Fayyaz Kashif[1], Linda Piscopo[1,2], Liam Collard[1,3], Cristian Ciracì[1], Massimo De Vittorio[1,2,3,*,†], Ferruccio Pisanello[1,3,*,†]

[1] *Istituto Italiano di Tecnologia, Center for Biomolecular Nanotechnologies, Arnesano, LE 73010, Italy*

[2] *Dipartimento di Ingegneria Dell'Innovazione, Università del Salento, Lecce 73100, Italy*

[3] *RAISE Ecosystem, Genova, Italy*

*Authors to whom correspondence should be addressed: di.zheng@iit.it, massimo.devittorio@iit.it, Ferruccio.pisanello@iit.it*

† *These authors contributed equally to this work*

**Abstract**

Creating plasmonic nanoparticles on a tapered optical fiber (TF) tip enables a remote SERS sensing probe, ideal for challenging sampling scenarios like biological tissue, specific cells, on-site environmental monitoring, and deep brain structures. However, nanoparticle patterns fabricated from current bottom-up methods are mostly random, making geometry control difficult. Uneven statistical distribution, clustering, and multilayer deposition introduce uncertainty in correlating device performance with morphology. Here, we employ a tunable solid-state dewetting method to create densely packed monolayer Au nanoislands (NIs) with varied geometric parameters, directly contacting the silica TF surface. These patterns exhibit analyzable nanoparticle sizes, densities, and uniform distribution across the entire taper surface, enabling a systematic investigation of particle size, density, and analyte effects on the SERS performance of the through-fiber detection system. The study is focused on the SERS response of a widely employed benchmark Rhodamine 6G (R6G) molecule and Serotonin, a neurotransmitter with high relevance for the neuroscience field. The numerical simulations and limit of detection (LOD) experiments on R6G show that the increase of the total near-field enhancement volume promotes the SERS sensitivity of the probe. However, for serotonin we observed a different behavior linked to its interaction with the nanoparticle's surface. The obtained LOD is as low as $10^{-7}$ M, a value not achieved so far in a through-fiber detection scheme. Therefore, we believe our work offers a strategy to design nanoparticle-based remote SERS sensing probes and provide new clues to discover and understand the intricate plasmonic-driven chemical reactions.

- *Introduction*

Molecular sensing based on surface-enhanced Raman scattering (SERS) has shown promise as a powerful analytical tool in various fields, including biomedical diagnostics, identification of chemical compounds, food safety, and environmental monitoring[1–4]. A frontier for SERS is represented by the ability to perform label-free and chemical-specific sensing in applications where sampling is difficult, including (i) probing biological tissue[5,6], (ii) site-specific study for living cells[7,8], and (iii) on-site environmental monitoring, such as detection for hazardous heavy metal and toxic gas[9–11]. Integrating a SERS substrate on a fiber optic probe can create a compact and multifunctional sensor that can be used as a point-of-care SERS amplifier or remote sensing probe where the excitation laser and the resulting Raman signal are guided and collected by the same waveguide (the through-fiber detection scheme). Additionally, the ability to implement SERS in small-diameter optical fibers[12,13] has the potential to extend SERS applications to environments where sample perturbations need to be minimized, such as brain structures. In the context of studying neurochemical dynamics, vibrational spectroscopy enhanced by plasmonic nanostructures is of particular interest by virtue of its ability to detect and identify neurotransmitters even at attomolar concentrations[14], with the release of these molecules linked to both physiological and pathological states, including Alzheimer's disease, depression, schizophrenia, and Parkinson's disease[15–18].

SERS-active fiber optic probes have been mainly fabricated by covering the termination of an optical fiber with metallic nanoparticles[19–31]. Of specific significance is the case of tapered optical fibers (TFs), which increase the surface area available for sensing but require the structuring of a curved surface with a non-constant curvature radius. Only bottom-up approaches have been demonstrated to successfully enable SERS sensing capabilities on a TF probe, including electrostatic self-assembly[10,27,32–39], direct nucleation reaction[40–43], dip coating, and laser-assisted evaporation[23,44–47]. However, plasmonic nanoparticle distributions on the surface fabricated by those approaches are largely random. The morphologies show an uneven statistical distribution across the large surface, cluster formation, and multilayer deposition, inevitably introducing uncertainty and ambiguity on the correlation between the device performance and the morphological status, limiting the understanding and design of the best-performance remote sensing tapered SERS probe.

Here, we aim at uncovering the correlation between particle size and density versus the probe's SERS performance, to provide design principles for particle-decorated TF remote sensing probes. A tunable solid-state dewetting technology was employed to fabricate nanoislands (NIs) with varied geometric parameters. The monolayer densely packed NIs are surfactant-free and in direct contact with the silica surface of TFs. The formed patterns have analyzable nanoparticle sizes, densities, and statistically uniform distributions through the entire extent of the taper's surface. This allows us to systematically study, for the first time, the effect of the particle size, density, and analyte on the overall SERS performance of through-fiber detection system.

By incrementally increasing the thickness of the deposited gold film, NIs patterns with gradually increased average diameters, heights, and gaps, were obtained. These increased NIs unit cells were numerically modeled to match with the statistically analyzed non-planar NIs patterns, finding that increasing NIs unit cells corresponds to increasing volume field enhancement (VFE) across the entire tapered fiber surface, thereby resulting in a gradual increase in SERS signals. Experimental evaluations were then performed with 785 nm excitation on four NIs-TFs configurations with gradually increased NIs unit cells, with rapid through-fiber SERS experiments that do not require a prolonged immersion in the analyte's solution. Firstly, the commonly used Raman reporter Rhodamine 6G (R6G) was used as the target analyte, due to its stable chemical composition and distinct Raman peaks. Across 4 different patterns of NIs-TFs probes, a stable limit of detection (LOD) of $10^{-6}$ M was obtained, while the spectral amplification gradients show that the increasing NIs unit cell facilitates SERS detection, aligning with the theoretical predictions on the VFE. Expanding the study to monoamine neurotransmitter, the use of serotonin as a target analyte revealed a different trend. The highest sensitivity (jointly considering the lowest LOD, number of detected peaks, and peak prominence) was indeed obtained with the second smallest NIs unit cells TF probes with a LOD of $10^{-7}$ M, which is, to the best of our knowledge, the lowest value to date in a through-fiber detection configuration. Unlike the neat and orderly spectral response of R6G, the serotonin spectral response shows sensitivity performance favoring small NIs unit cells. The notable attributes indicate that chemical reactions involving serotonin, driven by plasmonics, are occurring at the interface.

- ***Results***

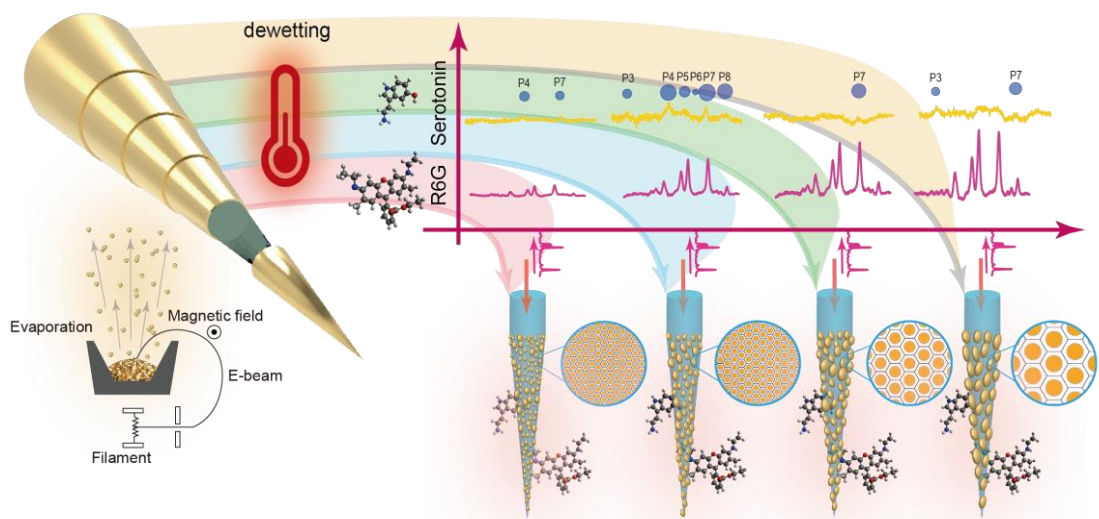

*Figure 1 A schematic illustration of the fabrication and characterization for tunable NIs-TF probes. The thicker the initial gold film deposition, the dewetting output generates bigger NIs diameters and gaps. The through-fiber SERS sensing for R6G and serotonin are analyzed from four different types of TFs with tunable NIs' patterns, the enlarged windows beside each tapered fiber illustrate the statistical particle size and density in the same scale.*

**Figure 1** shows schematic steps for fabricating NIs patterns with different geometrical parameters on the non-planar surface of the fiber taper. TFs were obtained by a heat-and-pull method from commercially available core/cladding step-index silica fibers

(core diameter 200 μm, NA 0.22)[48]. The tapered waveguide gradually reduces the number of guided photonic modes, and the guided and radiative modes can couple with plasmonic structures in direct contact with the taper surface[12,49]. During the pulling process, the high temperature (above the silica transition temperature of 1207 °C) results in a clean and smooth TF surface before evaporation. To obtain different film deposition thicknesses, we implemented the following procedure for each batch. 40 silica fibers were mounted on a rotational motor in the chamber of an e-beam evaporator, to ensure conformal deposition of a gold thin film over the entire TF's surface[50]. After the first layer of Au had been deposited to a nominal thickness of $thk_1 = 1.7$ nm, 10 TFs were removed from the chamber. Then a second layer of gold was deposited on top of the first layer on the remaining TFs with a thickness increased to $thk_2 = 3.3$ nm and 10 TFs were removed from the chamber. The procedure was repeated to obtain two more sets of TFs with film deposition thicknesses of $thk_3 = 5.0$ nm and $thk_4 = 6.7$ nm. Thermal annealing was then performed by raising the temperature from room temperature to 600 °C with a rate of 10 °C min$^{-1}$, after which the temperature was held at 600 °C for one hour. The thermal treatment was performed in standard atmospheric conditions, which has been reported to increase the bond between gold to fused silica by promoting diffusion of Au into the silica, due to the presence of oxygen[51,52]. Further details on the fabrication parameters are reported in Methods.

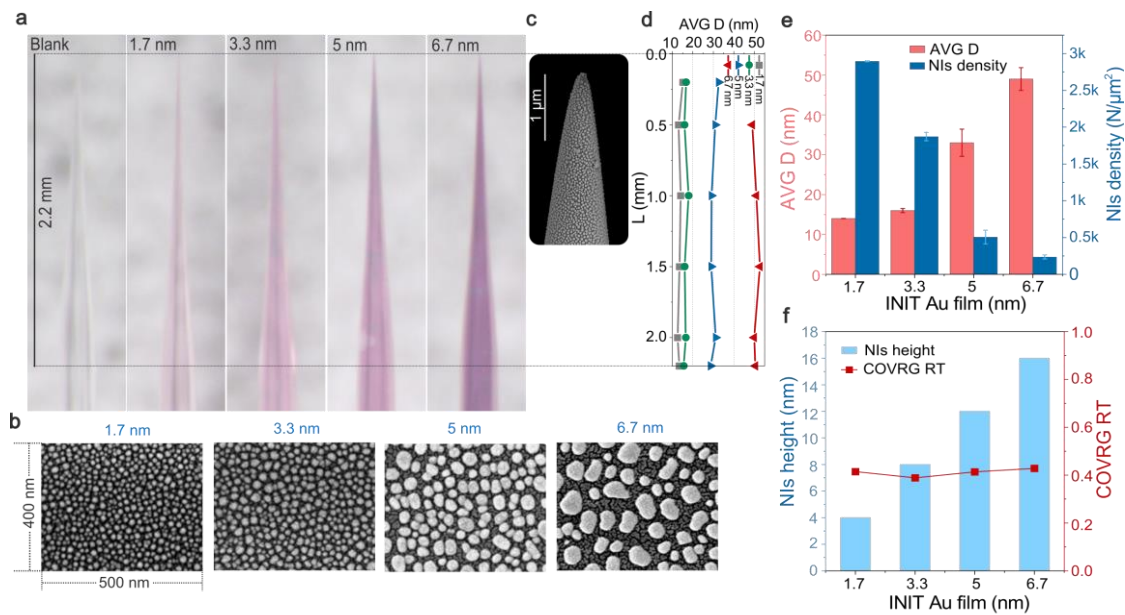

*Figure 2 Fabrication output for the tunable NIs-TFs set. (a) The optical images of a representative tunable NIs-TFs set, the blank TF on the left is used as color contrast reference, the following NIs-TFs from left to right are fabricated with initial Au film coating of 1.7, 3.3, 5, 6.7 nm respectively. (b) SEM images show the representative morphologies for NIs-TFs with initial Au film coating of 1.7, 3.3, 5, 6.7 nm respectively. (c) The SEM images of an enlarged view at the NIs-TF tip (Initial Au film of 5 nm). (d) The NIs average diameter distributions along the TF for a tunable NIs-TFs set. (e) The statistical analysis shows the average diameter (red bar) and the NIs density (blue bar) with the step increased initial Au film thickness. (f) The coverage rate (red square) and*

*NIs height (light blue bar) for each initial Au film thickness.*

The surface of the four sets of NIs-TFs at growing thickness shows a gradually increasing pink color under the optical microscope (**Fig.2a**), generated by an increase on NIs' size as identified by scanning electron microscope (SEM) (**Fig.2b**). To quantitatively extract the geometric parameters of the NIs patterns, SEM image batches were taken along the entire extent of the fiber taper (a representative high magnification image at the very tip of TF is shown in **Fig.2c**). Image recognition methods (tracing object boundaries and circular Hough transform) were implemented with custom code (see Methods) and were used to compute the coverage rate, particle number and particle density for each image. For the different Au film thicknesses, 2 TFs were included for the analysis. For each fiber, the number of sampled images with different magnifications along the taper was no less than 10. The measured average diameters at different taper positions are reported in **Fig.2d**, showing that the different NIs patterns distribute uniformly from the very tip to the taper end. The final geometric parameters of NIs' patterns were then extracted by averaging between two fibers, and each statistic is an averaging of all the data points at different positions along the whole taper. **Fig.2e** shows the average diameters and NIs' densities calculated, the diameters are $D_1 = 14 \pm 2$, $D_2 = 16 \pm 2$, $D_3 = 33 \pm 3$ and $D_4 = 49 \pm 3$ nm, and the NIs densities are $\rho_1 = 2895 \pm 7$, $\rho_2 = 1872 \pm 57$, $\rho_3 = 507 \pm 93$ and $\rho_4 = 236 \pm 27$ NIs/μm$^2$, for the i-th thk$_i$ (i = 1,2,3,4). **Fig.2f** displays the coverage rates ($C$) for the four different patterns, found to be about 40% independent of initial film thickness, and the corresponding average heights ($H_i$) of the NIs are $H_i$ = 4, 8, 12 and 16 nm (computed with $H_i = Thk_i / C_i$).

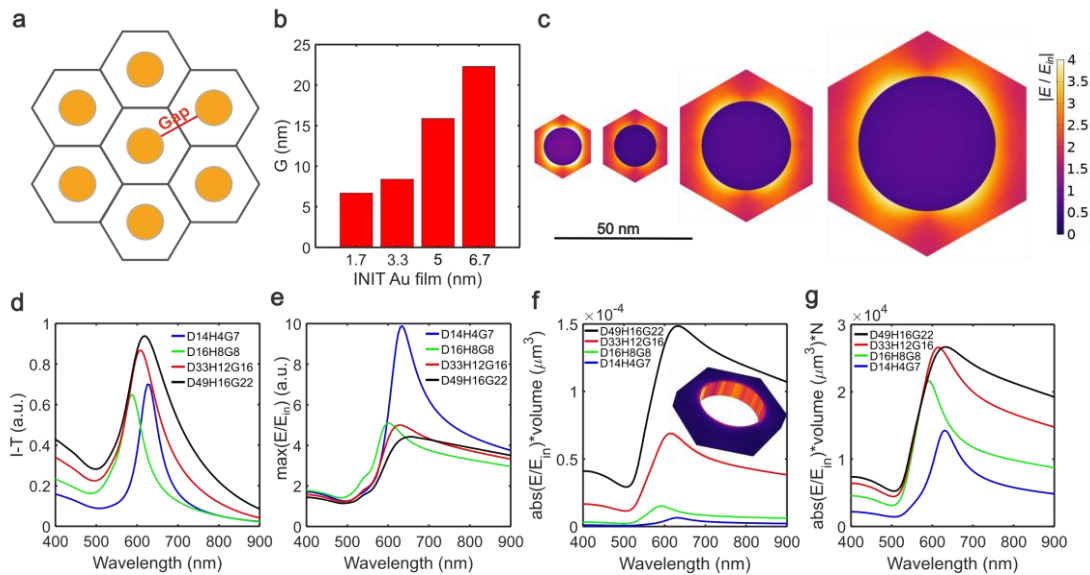

*Figure 3 Electromagnetic simulations based on SEM morphologies. (a) The illustration of periodic hexagonal NIs arrays with the red line indicating the gap. (b) The calculated gaps are according to hexagonal NIs patterns. (c) The near-field enhancement distribution for four morphologies and the relative sizes of four-unit cells correspond*

*with the real size ratio. (d) Plasmonic resonances are reported as 1-T for four morphologies of NIs patterns, where T is the calculated transmittance. (e) The maximum near-field enhancement along with different wavelengths produced by each morphology as a function of the incident wavelength under normal incidence. (f) The volumetric near-field enhancement is calculated by integrating the near-field enhancement over the hollow hexagonal prism surrounding the NIs as shown in the inset. (g) The volumetric near-field enhancement is multiplied by the number of unit cells that can occupy the whole taper fiber surface.*

To understand the electromagnetic performances of the different NIs patterns, we implemented a model with Au nanodisks arranged in a hexagonal periodic manner (**Fig.3a**). One nanodisk of diameter *D* occupies the center of a hexagonal unit cell. The diameters of the nanodisks are the average diameters for each pattern in **Fig.2e**. The heights of the nanodisks are taken from **Fig.2f** as $H_i = Thk_i / C_i$. The ratio between nanodisk's occupied surface and the overall unit cell area in 2D plane defines the coverage rates, which equal the coverage rates obtained from SEM image analysis (**Fig.2f**). Given the experimental average values for *D* and *C* as a function of *Thk*, the resulting effective gaps (*G*) in a hexagonal periodic arrangement can be computed to be 7, 8, 16 and 22 nm (**Fig.3b**), details in Methods. These geometric parameters define the NIs patterns and, hereinafter, we mark the patterns fabricated by initial film coating of $Thk_i$ = 1.7, 3.3, 5 and 6.7 nm as D14H4G7, D16H8G8, D33H12G16 and D49H16G22 respectively.

The electromagnetic behavior of the structures was simulated by a finite element method (commercial implementation by COMSOL Multiphysics) with an unpolarized plane wave excitation source. The spectral response was obtained by scanning the source's wavelength from 400 to 900 nm, while field enhancement maps were generated at 785 nm, a widely employed wavelength for Raman inspection of biological samples (details can be found in Methods). Representative field enhancement distributions in one unit cell are displayed in **Fig.3c**. The plasmonic resonances are shown as the complement of transmittance in **Fig.3d**: as each pattern has different geometric parameters, the plasmonic resonances of the systems are slightly different in terms of position, intensity, and spectral width. The maximum field enhancements as a function of wavelength are reported in **Fig.3e**, showing that at above 595 nm the smallest NIs D14H4G7 give the strongest field enhancement. Assuming that analytes are distributed uniformly in the surrounding environment, the absolute amount of the SERS signal for each unit cell correlates with the field enhancement integral in the space with non-zero field distribution (i.e. everywhere in the unit cell apart for the nanodisk). To better quantify the effect of electric field enhancement on the SERS performance, we calculated the volume field enhancement (VFE) by integrating the local filed enhancement $E/E_{in}$ over the volume of one unit cell, taken as an extruded hexagon around the nanodisk rising 10% over the nanodisks' height (inset in **Fig.3f,** cross-section in **Fig.S2**, VFE normalized to unit volume in **Fig.S3**). The calculated VFE shows that at all analyzed wavelengths, D49H16G22 has the strongest VFE, followed by D33H12G16, D16H8G8, and D14H4G7 (**Fig.3f**). Assuming that such patterns cover the entire fiber surface, we have estimated the number of unit cells *N* that covers the conical taper surface (see Methods). For a fixed surface area, the smaller NIs system will contain more unit cells than the bigger particles. **Fig.3g** reports the unit cell VFE multiplied by *N* for each pattern, showing that for most of the wavelength regions on which the calculations were performed the bigger particle patterns experience the higher absolute field enhancement (except for 570 to 620 nm). These results suggest that if the electromagnetic effect is the main contribution to the overall SERS signal, the larger particle pattern will generate the highest SERS scattered signal.

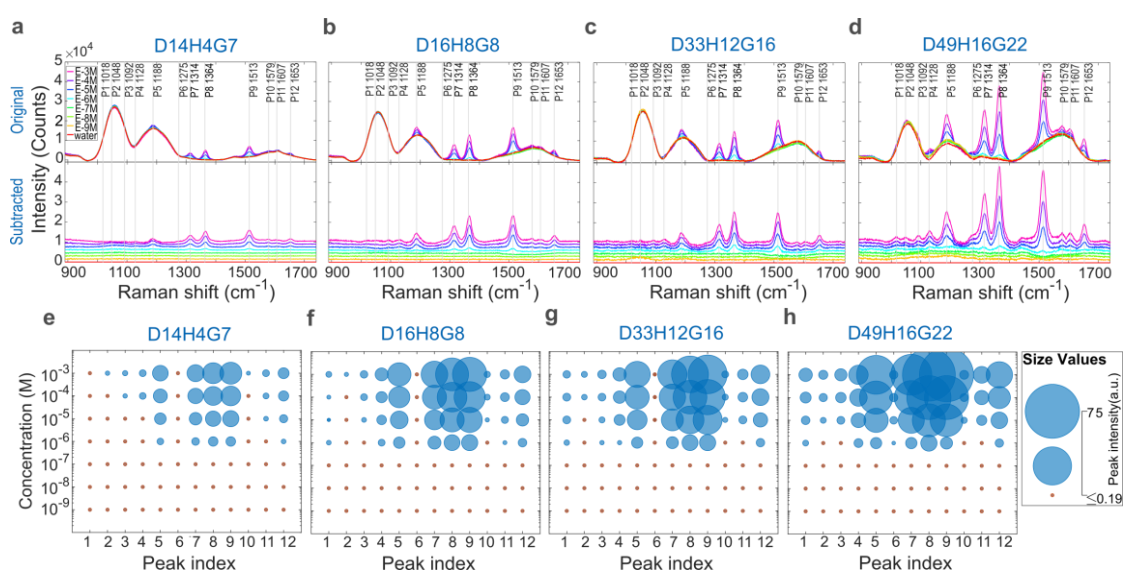

*Figure 4* Limit of detection experiment for aqueous solution of R6G with tunable NIs-TFs set. (a to d) The SERS spectra sets for tunable NIs-TFs with four NIs patterns (D14H4G7, D16H8G8, D33H12G16 and D49H16G22). The spectra in the top panels are the original spectra, and the spectra in the bottom panels are the silica background subtracted spectra (the silica background was taken as the spectra measured in water). The subtracted spectra have been vertically offset for clarity. The vertical lines mark the 12 R6G peak positions. (e to h) The bubble chart of 12 R6G peaks, the size of the blue bubbles corresponds to integrated peak areas, the small orange bubbles mark peak absence.

To verify the correlation of the NIs-TFs's performances with the 4 different NIs patterns on the TF surface, we conducted through-fiber SERS measurements in analytes solutions. We first used R6G as a chemically stable analyte to test the LOD of the probes. Different concentrations (from $10^{-9}$ to $10^{-3}$ M) of R6G aqueous solutions were prepared, then through-TF SERS measurements were conducted with a custom-built Raman microscope operating at 785 nm wavelength continuous laser excitation (details in Methods). For each NIs-TF, the measurements started from the lowest concentration solution and increased towards the highest concentration in steps, using a single fiber for each concentration ramp. Each spectrum was taken immediately after the fiber tip was immersed entirely into the aqueous solution, and the fiber tip was kept in the solution only during the exposure time. Four different patterns (D14H4G7, D16H8G8, D33H12G16 and D49H16G22) of NIs-TFs set were tested, with n = 2 fibers for each pattern. The measured spectra at different concentrations are reported in **Fig.4a-d**, in the original spectra, they are clearly composed by a contribution from the analyte and a Raman background generated by the probe itself. This latter is not the same for the four NIs patterns (Supplementary **Fig.S6**) and it has been subtracted to generate the analyte spectra reported in the bottom panels of **Fig.4a-d** (probe's background for subtraction was measured in water). For all types of NI-TFs, the R6G Raman signature can be identified from $10^{-6}$ M, and the peak intensities increase as the NIs grow. Among all the subtracted spectra, 12 R6G Raman peaks were identified (1018, 1048, 1092, 1128, 1188, 1275, 1314, 1364, 1513, 1579, 1607, 1653 cm$^{-1}$). In **Fig.4e-h**, a more detailed data analysis is presented with the bubble charts, allowing a better comparison between the different samples. The raw spectra were first normalized to the silica peak at 1055 cm$^{-1}$, then the silica background measured in water was subtracted from the normalized spectra. Working on the subtracted normalized spectra set, the 12 peaks were analyzed individually. A local linear baseline was removed from each peak, then we integrated the area of each peak. The summation of the corresponding peak intensities of the two data sets was used as a marker for the bubble sizes in the bubble chart (see supplementary **Fig.S5** for more details). For all 4 types of patterns, R6G signature peaks start to appear at the concentration of $10^{-6}$ M. Most of the peaks increase in intensity concurrently as concentration increases, and the patterns comprised of larger NI's, result in a more intense R6G SERS signal with a direct correlation to the VFE increase.

To verify if observations on R6G can be extended to other molecules with different chemical properties, we have applied the NI-TFs to the detection of serotonin, which is about 40% smaller than R6G and has a higher chemical affinity with gold. As an important monoamine neurotransmitter, serotonin regulates mood, sleep, appetite, cognition, pain sensation, and gastrointestinal functions, with imbalances linked to various conditions including depression, anxiety, insomnia, and cognitive impairments[53,54]. To understand the SERS sensing ability for tunable NIs-TF set on monoamine neurotransmitters, we performed LOD experiments with serotonin solutions. Again, tunable NIs-TFs sets with n = 2 fibers per pattern were measured. The representative spectra data sets and the corresponding subtracted spectra sets are shown in **Fig.5a-d**. The same data analysis approach as R6G was adopted (details in **Fig.S5**). There are 8 peaks identifiable across all the spectra data sets, at 908, 1036, 1115, 1339, 1420, 1482, 1538, and 1648 cm$^{-1}$. The detailed analysis results are shown in the bubble chart in **Fig.5e-h**. Interestingly, for serotonin, more peaks and lower LOD ($10^{-7}$ M) are obtained with smaller NIs patterns (D14H4G7 and D16H8G8). Additionally, from all the data sets there is an obvious degradation of the detection capability (or even vanishing) at high concentrations. Further tests on the probe that degraded at high serotonin concentrations show that it is still able to detect R6G molecules but with less peak intensity and one order of decreased detection sensitivity at $10^{-5}$ M (detailed data presented in **Fig.S6**). Considering the lowest LOD, the peak numbers and the spectra prominence, D16H4G8 NIs patterns show the best performance.

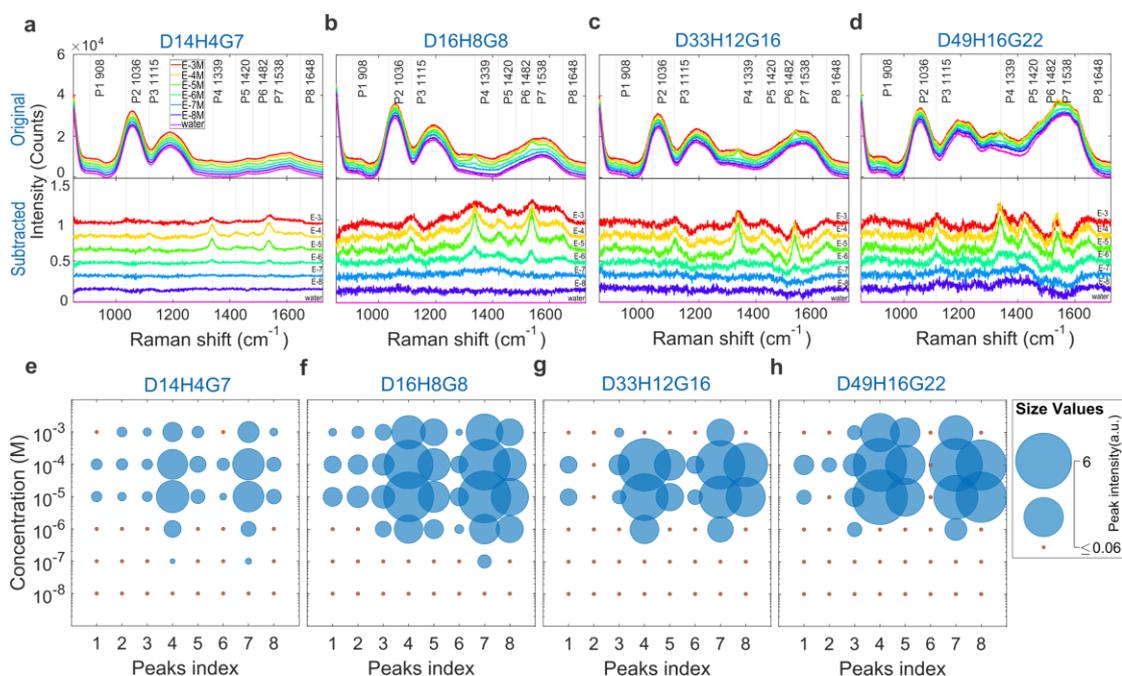

*Figure 5* *Limit of detection experiment for aqueous solutions of serotonin with tunable NIs-TFs set. (a to d) The SERS spectra sets for tunable NIs-TFs with four NIs patterns (D14H4G7, D16H8G8, D33H12G16 and D49H16G22). The spectra in the top panels are the original spectra, and the spectra in the bottom panels are the silica background subtracted spectra (the silica background was taken as the spectra measured in water), The spectra sets have been vertically offset for clarity. The vertical lines mark the 8 serotonin peak positions. (e to h) The bubble chart of 8 serotonin peaks, the size of the blue bubbles corresponds to integrated peak areas, the small orange bubbles mark peak absence.*

Comparing the spectra responses of R6G and serotonin, there are both common aspects as well as differences. The common aspects mainly relate to the probe's background behavior, which is summarized in Supplementary **Fig.S7**. The peak at 1055 $cm^{-1}$ is relatively constant, while the signal at higher wavenumbers (above 1100 $cm^{-1}$) increases with the increase of the NIs' size. This shows that the different NIs patterns on the tapered tip can do wavelength selective modulation on the ratio between reflection and transmission of the total generated Raman signal (both the silica waveguide Raman and molecular SERS contributions). In the case of differences instead, they can be listed as follows: (i) Within one set of spectra collected from the same fiber, when the concentration increases, the peaks' intensities of R6G increase, while for serotonin the peaks' intensities start to decrease at $10^{-4}$ M, and this degradation of peaks is stronger at increasing concentrations; (ii) For R6G more peaks are appearing along with increased total VFE, while for serotonin this happens for samples with nominally lower VFE (D14H4G7 and D16H8G8). (iii) The smaller NIs with narrower gaps (D16H8G8) can achieve a lower LOD for serotonin with no main changes in peak intensity compared with higher VFE patterns, while for R6G peaks prominence is increased along with increased VFE.

- **Discussion and conclusion**

In a through-fiber detection configuration with plasmonic NIs-TFs, several factors can influence the ability to detect SERS signals from a target analyte. These include: (1) The ability of NIs patterns to generate near-field enhancement, (2) the relative strength between molecular SERS signal and the Raman background generated by the fiber when the excitation laser goes through, (3) the interaction between analytes and gold NIs, in terms of potential chemical reactions on the surface, and (4) the spatial overlap between analytes and the near-field enhancement. In our study, we have observed that these factors differently influence the commonly employed R6G molecule used to benchmark SERS performances and the neurologically relevant neurotransmitter serotonin.

For R6G our experiments show that total VFE of the tunable NIs patterns on the taper surface mainly dominates the spectral response. The higher SERS signals can be obtained along with gradually increased total VFE. The modulation on the background reflection does not affect the intensity of characteristic peaks significantly, since R6G has a big Raman cross-section which can provide high relative strength when mixing with the silica Raman background from the fiber. R6G does not possess functional groups with a strong affinity to gold (therefore it is unlikely that R6G will bond on NIs and generate surface chemical reactions), and the spectral response is stable with well-recognized molecular vibrations[55,56]. Additionally, the molecular size of R6G (~1 nm[57,58]) is well below the narrowest investigated gap (~7nm). This latter consideration, along with the low affinity with gold, let us hypothesize that the spatial overlap between R6G molecules and the near-field enhancement can be considered uniform, matching well with the assumption made in electromagnetic simulations.

For the more complex spectral response of serotonin, its small Raman cross-section results in a weak relative strength when mixing with the fiber's Raman signal, which makes the modulation on the background have more impact on the detection of serotonin peaks. Thus, the increasing background observed as a function of the NI's size is more likely to diminish the peaks' prominence at low concentrations. This could be one reason explaining the lower LOD obtained for samples D14H4G7 and D16H8G8 compared to D33H12G16 and D49H16G22. Also, the smaller serotonin molecules (roughly 60% of R6G in size) with the amino functional group have been reported able to bond on gold surfaces[59–63]. Therefore, serotonin molecules have a better chance to bond at the very narrow gap region and the very vicinity of the gold surface even when concentrations are low, resulting in better spatial overlap between molecules and more intense 'hot spots' generated by the narrower gaps of the smaller NIs patterns (D14H4G7 and D16H8G8). Along with concentration increase, the increased density of serotonin bonded on the gold surface might induce a polymerization reaction[64] forming a Raman inactive dielectric layer dampening the SERS sensitivity for serotonin molecules at high concentrations ($\geq 10^{-4}$ M). The interplay of all those factors results in the serotonin spectra response shown in **Fig.5**. It is important to highlight that the detected serotonin SERS bands are slightly different in the available literature reports, in terms of both the obtained number of peaks and their positions, as affected by the excitation wavelength, dissolving environments and employed nanometal materials[14,65–67]. The peak disappearance and the downshifts of peaks associated with in-phase breathing and stretching modes of the indole ring were also reported [68]. For more informative studies about surface chemical reactions on metal-bound reactants, the focus is mainly on thiol molecules[69–72], which are well known to be able to strongly bond with Au surfaces. In a nanoparticle-molecule-gold film junction, 4-nitrobenzenethiol (NBT) has been reported dimerized into 4,4′-dimercaptoazobenzene (DMAB) under laser radiation, and the SERS spectrum shows multiple new peaks after reaction[69]. A similar phenomenon has also been reported in an Au nanoparticle on microsphere system[72].

In summary, within the bottom-up methods to decorate nanoparticles on the tapered fiber tip, the tunable solid-state dewetting technology offers an effective and robust way to fabricate surfactant-free, well-defined, uniformly distributed, and densely packed monolayer NIs with controllable geometric parameters. Together with potential applications in remote detection schemes, the unique geometrical tunability helps in better understanding the correlation between particle size and density versus the electromagnetic performance of the probes, showing the electromagnetic part of the SERS performance of the probe can be optimized by maximizing the total VFE, other than the maximum field enhancement within the gaps. With the analyzable electromagnetic performance of the probe, the potential plasmonic-driven chemical reactions concerning monoamine neurotransmitters can be recognized. Thus, we believe that our work introduces a broad-reaching methodology for designing remote SERS sensing probes based on nanoparticles, along with providing an experimental paradigm for novel insights into the intricate surface interactions propelled by plasmonic resonances.

- **Methods**

**Tunable NIs-TFs set fabrication.**
The tapered fibers (TFs) were fabricated from the standard multimode silica optical fibers (FG200LEA, 0.22 NA, Low-OH, Ø200 μm Core). To start with, the silica fibers were cut to a fixed length of 12 cm, then all the fibers were put in an acetone bath for 30 mins to remove the acrylate jacket. Using the puller system (Sutter P-2000), the silica fibers were pulled rapidly, after laser heating, with very small taper angles. After the silica TFs were prepared, all the TFs were sent to an e-beam evaporator for Au thin film deposition (Thermionics Laboratory, inc. e-GunTM). During the evaporation, all the TFs were mounted on a rotational motor to ensure conformal gold evaporation over the entire TFs' surfaces. The evaporation rate of 0.2 Å/s was used for all the depositions, with chamber pressure $< 8 \times 10^{-7}$ mbar. For a batch of the tunable NIs-TFs set fabrication, around 40 blank silica fibers were placed in the chamber for the thin film deposition. Once the first layer of gold with a nominal thickness of thk1 = 1.7 nm was deposited, 10 thin film-deposited TFs were extracted from the chamber. Subsequently, a second layer of gold was deposited on top of the remaining TFs, contributing to a combined thickness of thk2 = 3.3 nm. After this deposition, another 10 TFs were withdrawn from the chamber. This process was iterated to generate two additional batches of TFs, each with deposition thicknesses of thk3 = 5.0 nm and thk4 = 6.7 nm, respectively. For the film thickness control, the TFs were linked to flat surface evaporation with relation $T_{TFs} = 1/\pi \times T_{Flat}$, thus, for Thk = 1.7, 3.3, 5, and 6.7 nm gold film deposition on TFs, the film sensor of the evaporator monitors 5, 10, 15, 20 nm of Au deposition. When the evaporation procedure finished, the fibers were detached from the mount and arranged in a ceramic bowl without any adhesive for thermal annealing in a muffle furnace (Nabertherm B180). The furnace was set to gradually ramp up from room temperature (RT) to 600 °C with a rate of 10 °C min$^{-1}$, and held at 600 °C for 1h, then allowed to cool ambiently to RT. Eventually, the non-structured side of fabricated fibers was connected to metallic ferrules with a diameter of 1.25 mm, and went through a manual polishing process.

**The morphological analysis of tunable NIs patterns.**
For the morphological analysis, SEM images were acquired with the FEI Helios Nanolab 600i Dual Beam system. The SEM image acquisitions were conducted after sputtering a thin layer of gold on all the NIs surfaces to provide sufficient conductivity. The sampling of the images was done through the entire extent of the fiber, and the number of samplings is no less than 10. Home-developed MATLAB programs were used for morphological analysis. The NIs number count and the coverage situations are obtained by a combined method of tracing object boundaries and circular Hough transform. Eventually, the coverage rates (*C*) and the average areas of physical occupation (*S*) were extracted, thus the average diameters can be determined by $D = 2 \times \sqrt{S/\pi}$. The average height *(H)* of the NIs can be computed by $H = Thk/C$, where *Thk* is the initial film deposition thickness. We have included section S1 in the supplementary for a detailed description, and the analysis program is available on reasonable request.

**Electromagnetic simulations.**
The geometric parameters obtained from morphological analysis, including average diameters (*D*), height (*H*), and coverage rate (*C*), were used to build up a simplified numerical model with COMSOL Multiphysics, to emphasize the effect of the interparticle distances and reduce the computational cost. Thus, each pattern is represented with Au disks arranged in a periodic hexagonal array, with Au disks (all with diameter *D*) in the center of the hexagonal unit cell. To mimic the droplet-like shape of the fabricated sample, we have rounded the top edge of the disk with a radius of curvature $r = H \times 10\%$. In the 2D plane of the hexagonal unit cell, *C* is defined as the ratio between the gold disk area and the hexagonal unit cell area. In this picture, we can deduce the effective gap (*G*) as:

$$G = D \times \left(\sqrt{\frac{\sqrt{3}\pi}{6C}} - 1\right)$$

By simulating three-dimensional time-harmonic Maxwell's equations, we obtained the results in Fig.3. The particle sits on an infinite dielectric substrate with a refractive index n = 1.4. Optical constants of gold were obtained from Ref.[73]. The mesh resolution is chosen so that transmittance results are converged to < 0.1% at resonance.

To determine the number of unit cells (N) occupying the whole TF surface for 4 patterns, the tapered fibers were considered as a right circular cone with the diameter of its base $d$ = 220 μm, its height $h$ = 2.2 mm, thus we can calculate the surface area $s = \frac{\pi d}{2} \times \sqrt{h^2 + \left(\frac{d}{2}\right)^2}$ = 761215 μm². The NIs densities are 2895, 1872, 507, and 236 NIs/μm² for four patterns obtained from SEM image analysis in **Fig2e**, thus N = [2.2037, 1.4250, 0.3859, 0.1796] × $10^9$.

**Through-fiber SERS characterizations.**

The optical setup for the home-built Raman microscope is shown in **Fig.S4**. briefly, the linear polarized free space laser of 785 nm continuous wavelength was coupled into a meter-long fiber patch cord to launch the laser into the excitation path of the Raman microscope, the resulting excitation laser delivered to the sample was scrambled into a speckle pattern.

The collimated laser beam filled the back aperture of the focus lens L1 (aspheric, Ø25.0 mm, f = 20 mm), resulting in a light spot of 200 μm in diameter, which matches the fiber core size. The fibers were configured with the distal end facing the focus lens, the laser was injected over the full angular acceptance of the fiber (NA 0.22) to recruit most of the propagating modes. The Raman signal, collected by the TF, was then separated from the pump laser using a dichroic mirror (DC: Semrock, LPD02-785RU-25) and a long-pass razor-edge filter (F1: Semrock, LP02-785RU-25). Then, the signal was routed to a spectrometer (Horiba iHR320). The Raman measurements were performed with a slit at 200 μm and a 600 l/mm (blaze 750 nm) grating. Spectra were recorded on a SYNAPSE CCD cooled to -50 °C. All the raw spectra were treated with baseline correction (ALS). In the through-TFs SERS measurements. The laser power used for all measurements is 68 mW, the spectra acquisition time ranges between 30 s to 180 s, and for analyzed data sets, all spectra intensity were normalized to 60 s exposure time. R6G (Rhodamine 6G, $C_{28}H_{31}N_2O_3Cl$), dopamine (3-hydroxytyramine hydrochloride, $C_8H_{12}ClNO_2$), and serotonin (5-hydroxytryptamine, $C_{10}H_{12}N_2O$) were all purchased from Merck KGaA.

**Supporting Information**

Supporting Information is available.

**Acknowledgments**

M.D.V. and F.P. jointly supervised and are co-last authors in this work.

D.Z., M.F.K., L.P., Li.C., C.C., M.D.V., and F.P. acknowledge funding from the European Union's Horizon 2020 Research and Innovation Program under Grant Agreement No. 828972. L.C., M.D.V. and F.P. acknowledge funding from the Project "RAISE (Robotics and AI for Socio-economic Empowerment)" code ECS00000035 funded by European Union – NextGenerationEU PNRR MUR - M4C2 – Investimento 1.5 - Avviso "Ecosistemi dell'Innovazione" CUP J33C22001220001. F.P. acknowledges funding from the European Research Council under the European Union's Horizon 2020 Research and Innovation Program under Grant Agreement No. 677683. M.D.V., and F.P. acknowledge that this project has received funding from the European Union's Horizon 2020 Research and Innovation Program under Grant Agreement No. 101016787. F.P., and M.D.V. were funded by the U.S. National Institutes of Health (Grant No. 1UF1NS108177-01).

**Reference**

(1) Choi, N.; Schlücker, S. Convergence of Surface-Enhanced Raman Scattering with Molecular Diagnostics: A Perspective on Future Directions. *ACS Nano* **2024**, *18* (8), 5998–6007. https://doi.org/10.1021/acsnano.3c11370.
(2) Chen, C.; Wang, X.; Wang, R.; Waterhouse, G. I. N.; Xu, Z. SERS-Tag Technology in Food Safety and Detection: Sensing from the "Fingerprint" Region to the "Biological-Silent" Region. *Journal of Future Foods* **2024**, *4* (4), 309–323. https://doi.org/10.1016/j.jfutfo.2023.11.003.
(3) Vázquez-Iglesias, L.; Stanfoca Casagrande, G. M.; García-Lojo, D.; Ferro Leal, L.; Ngo, T. A.; Pérez-Juste, J.; Reis, R. M.; Kant, K.; Pastoriza-Santos, I. SERS Sensing for Cancer Biomarker: Approaches and Directions. *Bioactive Materials* **2024**, *34*, 248–268. https://doi.org/10.1016/j.bioactmat.2023.12.018.
(4) Ong, T. T. X.; Blanch, E. W.; Jones, O. A. H. Surface Enhanced Raman Spectroscopy in Environmental Analysis, Monitoring and Assessment. *Science of The Total Environment* **2020**, *720*, 137601. https://doi.org/10.1016/j.scitotenv.2020.137601.
(5) Bergholt, M. S.; Zheng, W.; Ho, K. Y.; Teh, M.; Yeoh, K. G.; So, J. B. Y.; Shabbir, A.; Huang, Z. Fiber-Optic Raman Spectroscopy Probes Gastric Carcinogenesis in Vivo at Endoscopy. *Journal of Biophotonics* **2013**, *6* (1), 49–59. https://doi.org/10.1002/jbio.201200138.
(6) McGregor, H. C.; Short, M. A.; Lam, S.; Shaipanich, T.; Beaudoin, E.-L.; Zeng, H. Development and in Vivo Test of a Miniature Raman Probe for Early Cancer Detection in the Peripheral Lung. *Journal of Biophotonics* **2018**, *11* (11), e201800055. https://doi.org/10.1002/jbio.201800055.
(7) Fortuni, B.; Ricci, M.; Vitale, R.; Inose, T.; Zhang, Q.; Hutchison, J. A.; Hirai, K.; Fujita, Y.; Toyouchi, S.; Krzyzowska, S.; Van Zundert, I.; Rocha, S.; Uji-i, H. SERS Endoscopy for Monitoring Intracellular Drug Dynamics. *ACS Sens.* **2023**, *8* (6), 2340–2347. https://doi.org/10.1021/acssensors.3c00394.
(8) Zhang, Q.; Inose, T.; Ricci, M.; Li, J.; Tian, Y.; Wen, H.; Toyouchi, S.; Fron, E.; Ngoc Dao, A. T.; Kasai, H.; Rocha, S.; Hirai, K.; Fortuni, B.; Uji-i, H. Gold-Photodeposited Silver Nanowire Endoscopy for Cytosolic and Nuclear pH Sensing. *ACS Appl. Nano Mater.* **2021**, *4* (9), 9886–9894. https://doi.org/10.1021/acsanm.1c02363.
(9) Zhang, H.; Zhou, X.; Li, X.; Gong, P.; Zhang, Y.; Zhao, Y. Recent Advancements of LSPR

Fiber-Optic Biosensing: Combination Methods, Structure, and Prospects. *Biosensors* **2023**, *13* (3), 405. https://doi.org/10.3390/bios13030405.

(10) Liu, Y.; Lin, C.; Chen, H.; Shen, C.; Zheng, Z.; Li, M.; Xu, B.; Zhao, C.; Kang, J.; Wang, Y. A Rapid Surface-Enhanced Raman Scattering Method for the Determination of Trace $Hg^{2+}$ with Tapered Optical Fiber Probe. *Microchemical Journal* **2024**, *196*, 109724. https://doi.org/10.1016/j.microc.2023.109724.

(11) Lyu, D.; Huang, Q.; Wu, X.; Nie, Y.; Yang, M. Optical Fiber Sensors for Water and Air Quality Monitoring: A Review. *OE* **2023**, *63* (3), 031004. https://doi.org/10.1117/1.OE.63.3.031004.

(12) Zheng, D.; Pisano, F.; Collard, L.; Balena, A.; Pisanello, M.; Spagnolo, B.; Mach-Batlle, R.; Tantussi, F.; Carbone, L.; De Angelis, F.; Valiente, M.; de la Prida, L. M.; Ciracì, C.; De Vittorio, M.; Pisanello, F. Toward Plasmonic Neural Probes: SERS Detection of Neurotransmitters through Gold-Nanoislands-Decorated Tapered Optical Fibers with Sub-10 Nm Gaps. *Advanced Materials* **2023**, *35* (11), 2200902. https://doi.org/10.1002/adma.202200902.

(13) Collard, L.; Pisano, F.; Zheng, D.; Balena, A.; Kashif, M. F.; Pisanello, M.; D'Orazio, A.; de la Prida, L. M.; Ciracì, C.; Grande, M.; De Vittorio, M.; Pisanello, F. Holographic Manipulation of Nanostructured Fiber Optics Enables Spatially-Resolved, Reconfigurable Optical Control of Plasmonic Local Field Enhancement and SERS. *Small* **2022**, *18* (23), 2200975. https://doi.org/10.1002/smll.202200975.

(14) Lee, W.; Kang, B.-H.; Yang, H.; Park, M.; Kwak, J. H.; Chung, T.; Jeong, Y.; Kim, B. K.; Jeong, K.-H. Spread Spectrum SERS Allows Label-Free Detection of Attomolar Neurotransmitters. *Nat Commun* **2021**, *12* (1), 159. https://doi.org/10.1038/s41467-020-20413-8.

(15) Malhi, G. S.; Mann, J. J. Depression. *The Lancet* **2018**, *392* (10161), 2299–2312. https://doi.org/10.1016/S0140-6736(18)31948-2.

(16) McCutcheon, R. A.; Krystal, J. H.; Howes, O. D. Dopamine and Glutamate in Schizophrenia: Biology, Symptoms and Treatment. *World Psychiatry* **2020**, *19* (1), 15–33. https://doi.org/10.1002/wps.20693.

(17) Segura-Aguilar, J.; Paris, I.; Muñoz, P.; Ferrari, E.; Zecca, L.; Zucca, F. A. Protective and Toxic Roles of Dopamine in Parkinson's Disease. *Journal of Neurochemistry* **2014**, *129* (6), 898–915. https://doi.org/10.1111/jnc.12686.

(18) Xu, Y.; Yan, J.; Zhou, P.; Li, J.; Gao, H.; Xia, Y.; Wang, Q. Neurotransmitter Receptors and Cognitive Dysfunction in Alzheimer's Disease and Parkinson's Disease. *Progress in Neurobiology* **2012**, *97* (1), 1–13. https://doi.org/10.1016/j.pneurobio.2012.02.002.

(19) Du, W.; Wei, S.; Li, N.; Hao, Z.; Li, Y.; Wang, M. Highly Sensitive Fiber Optic Enhanced Raman Scattering Sensor. *Optics & Laser Technology* **2024**, *168*, 109879. https://doi.org/10.1016/j.optlastec.2023.109879.

(20) Huang, R.; Lian, S.; Li, J.; Feng, Y.; Bai, S.; Wu, T.; Ruan, M.; Wu, P.; Li, X.; Cai, S.; Jiang, P. High-Sensitivity and Throughput Optical Fiber SERS Probes Based on Laser-Induced Fractional Reaction Method. *Results in Physics* **2023**, *48*, 106410. https://doi.org/10.1016/j.rinp.2023.106410.

(21) Yang, X.; Ileri, N.; Larson, C. C.; Carlson, T. C.; Britten, J. A.; Chang, A. S. P.; Gu, C.; Bond, T. C. Nanopillar Array on a Fiber Facet for Highly Sensitive Surface-Enhanced Raman Scattering. *Opt. Express, OE* **2012**, *20* (22), 24819–24826.

https://doi.org/10.1364/OE.20.024819.

(22) Huang, J.; Zhou, F.; Cai, C.; Chu, R.; Zhang, Z.; Liu, Y. Remote SERS Detection at a 10-m Scale Using Silica Fiber SERS Probes Coupled with a Convolutional Neural Network. *Opt. Lett.* **2023**, *48* (4), 896. https://doi.org/10.1364/OL.483939.

(23) Yu, Z.; Wang, Z.; Zhang, J. Preparation Optimization for a Silver Cavity Coupled Tapered Fiber SERS Probe with High Sensitivity. *Opt. Mater. Express, OME* **2022**, *12* (7), 2835–2843. https://doi.org/10.1364/OME.459758.

(24) Tian, Q.; Cao, S.; He, G.; Long, Y.; Zhou, X.; Zhang, J.; Xie, J.; Zhao, X. Plasmonic Au-Ag Alloy Nanostars Based High Sensitivity Surface Enhanced Raman Spectroscopy Fiber Probes. *Journal of Alloys and Compounds* **2022**, *900*, 163345. https://doi.org/10.1016/j.jallcom.2021.163345.

(25) Kang, T.; Cho, Y.; Yuk, K. M.; Yu, C. Y.; Choi, S. H.; Byun, K. M. Fabrication and Characterization of Novel Silk Fiber-Optic SERS Sensor with Uniform Assembly of Gold Nanoparticles. *Sensors* **2022**, *22* (22), 9012. https://doi.org/10.3390/s22229012.

(26) He, G.; Han, X.; Cao, S.; Cui, K.; Tian, Q.; Zhang, J. Long Spiky Au-Ag Nanostar Based Fiber Probe for Surface Enhanced Raman Spectroscopy. *Materials* **2022**, *15* (4), 1498. https://doi.org/10.3390/ma15041498.

(27) Yu, M.; Tian, Q.; He, G.; Cui, K.; Zhang, J. Surface-Enhanced Raman Scattering Fiber Probe Based on Silver Nanocubes. *Adv. Fiber Mater.* **2021**, *3* (6), 349–358. https://doi.org/10.1007/s42765-021-00106-7.

(28) Zhou, F.; Liu, Y.; Wang, H.; Wei, Y.; Zhang, G.; Ye, H.; Chen, M.; Ling, D. Au-Nanorod-Clusters Patterned Optical Fiber SERS Probes Fabricated by Laser-Induced Evaporation Self-Assembly Method. *Opt. Express, OE* **2020**, *28* (5), 6648–6662. https://doi.org/10.1364/OE.386215.

(29) Kim, J. A.; Wales, D. J.; Thompson, A. J.; Yang, G.-Z. Fiber-Optic SERS Probes Fabricated Using Two-Photon Polymerization For Rapid Detection of Bacteria. *Advanced Optical Materials* **2020**, *8* (9), 1901934. https://doi.org/10.1002/adom.201901934.

(30) Kwak, J.; Lee, W.; Kim, J.-B.; Bae, S.-I.; Jeong, K.-H. Fiber-Optic Plasmonic Probe with Nanogap-Rich Au Nanoislands for on-Site Surface-Enhanced Raman Spectroscopy Using Repeated Solid-State Dewetting. *JBO* **2019**, *24* (3), 037001. https://doi.org/10.1117/1.JBO.24.3.037001.

(31) Xia, M.; Zhang, P.; Leung, C.; Xie, Y.-H. SERS Optical Fiber Probe with Plasmonic End-Facet. *Journal of Raman Spectroscopy* **2017**, *48* (2), 211–216. https://doi.org/10.1002/jrs.5031.

(32) Qin, Y.; Huang, R.; Lu, F.; Tang, H.; Yao, B.; Mao, Q. Effects of the Cone Angle on the SERS Detection Sensitivity of Tapered Fiber Probes. *Opt. Express, OE* **2022**, *30* (21), 37507–37518. https://doi.org/10.1364/OE.471597.

(33) Li, T.; Yu, Z.; Wang, Z.; Zhu, Y.; Zhang, J. Optimized Tapered Fiber Decorated by Ag Nanoparticles for Raman Measurement with High Sensitivity. *Sensors* **2021**, *21* (7), 2300. https://doi.org/10.3390/s21072300.

(34) Zhu, H.; Masson, J.-F.; Bazuin, C. G. Templating Gold Nanoparticles on Nanofibers Coated with a Block Copolymer Brush for Nanosensor Applications. *ACS Appl. Nano Mater.* **2020**, *3* (1), 516–529. https://doi.org/10.1021/acsanm.9b02081.

(35) Hutter, T.; Elliott, S. R.; Mahajan, S. Optical Fibre-Tip Probes for SERS: Numerical Study for

Design Considerations. *Opt. Express* **2018**, *26* (12), 15539. https://doi.org/10.1364/OE.26.015539.

(36) Xu, W.; Chen, Z.; Chen, N.; Zhang, H.; Liu, S.; Hu, X.; Wen, J.; Wang, T. SERS Taper-Fiber Nanoprobe Modified by Gold Nanoparticles Wrapped with Ultrathin Alumina Film by Atomic Layer Deposition. *Sensors* **2017**, *17* (3), 467. https://doi.org/10.3390/s17030467.

(37) Lussier, F.; Brulé, T.; Bourque, M.-J.; Ducrot, C.; Trudeau, L.-É.; Masson, J.-F. Dynamic SERS Nanosensor for Neurotransmitter Sensing near Neurons. *Faraday Discuss.* **2017**, *205*, 387–407. https://doi.org/10.1039/C7FD00131B.

(38) Huang, Z.; Lei, X.; Liu, Y.; Wang, Z.; Wang, X.; Wang, Z.; Mao, Q.; Meng, G. Tapered Optical Fiber Probe Assembled with Plasmonic Nanostructures for Surface-Enhanced Raman Scattering Application. *ACS Appl. Mater. Interfaces* **2015**, *7* (31), 17247–17254. https://doi.org/10.1021/acsami.5b04202.

(39) Chen, Z.; Dai, Z.; Chen, N.; Liu, S.; Pang, F.; Lu, B.; Wang, T. Gold Nanoparticles-Modified Tapered Fiber Nanoprobe for Remote SERS Detection. *IEEE Photonics Technology Letters* **2014**, *26* (8), 777–780. https://doi.org/10.1109/LPT.2014.2306134.

(40) Cao, J.; Zhao, D.; Qin, Y. Novel Strategy for Fabrication of Sensing Layer on Thiol-Functionalized Fiber-Optic Tapers and Their Application as SERS Probes. *Talanta* **2019**, *194*, 895–902. https://doi.org/10.1016/j.talanta.2018.11.012.

(41) Cao, J.; Zhao, D.; Mao, Q. A Highly Reproducible and Sensitive Fiber SERS Probe Fabricated by Direct Synthesis of Closely Packed AgNPs on the Silanized Fiber Taper. *Analyst* **2017**, *142* (4), 596–602. https://doi.org/10.1039/C6AN02414A.

(42) Zhang, J.; Chen, S.; Gong, T.; Zhang, X.; Zhu, Y. Tapered Fiber Probe Modified by Ag Nanoparticles for SERS Detection. *Plasmonics* **2016**, *11* (3), 743–751. https://doi.org/10.1007/s11468-015-0105-1.

(43) Cao, J.; Zhao, D.; Mao, Q. Laser-Induced Synthesis of Ag Nanoparticles on the Silanized Surface of a Fiber Taper and Applications as a SERS Probe. *RSC Adv.* **2015**, *5* (120), 99491–99497. https://doi.org/10.1039/C5RA18504A.

(44) Volkan, M.; Stokes, D. L.; Vo-Dinh, T. Surface-Enhanced Raman of Dopamine and Neurotransmitters Using Sol-Gel Substrates and Polymer-Coated Fiber-Optic Probes. *Appl Spectrosc* **2000**, *54* (12), 1842–1848. https://doi.org/10.1366/0003702001948952.

(45) Liu, Y.; Liu, R.; Ai, C.; Wang, B.; Chu, R.; Wang, H.; Shui, L.; Zhou, F. Stick-Slip-Motion-Assisted Interfacial Self-Assembly of Noble Metal Nanoparticles on Tapered Optical Fiber Surface and Its Application in SERS Detection. *Applied Surface Science* **2022**, *602*, 154298. https://doi.org/10.1016/j.apsusc.2022.154298.

(46) Kaur, N.; Das, G. Three-Dimensional Plasmonic Substrate as Surface-Enhanced Raman Spectroscopy (SERS) Tool for the Detection of Trace Chemicals. *Journal of Raman Spectroscopy* *n/a* (n/a). https://doi.org/10.1002/jrs.6649.

(47) Tao, P.; Ge, K.; Dai, X.; Xue, D.; Luo, Y.; Dai, S.; Xu, T.; Jiang, T.; Zhang, P. Fiber Optic SERS Sensor with Silver Nanocubes Attached Based on Evanescent Wave for Detecting Pesticide Residues. *ACS Appl. Mater. Interfaces* **2023**, *15* (25), 30998–31008. https://doi.org/10.1021/acsami.3c04059.

(48) Sileo, L.; Pisanello, M.; De Vittorio, M.; Pisanello, F. Fabrication of Multipoint Light Emitting Optical Fibers for Optogenetics. *Proc. SPIE* **2015**, *9305*, 93052O. https://doi.org/10.1117/12.2075819.

(49) Pisano, F.; Kashif, M. F.; Balena, A.; Pisanello, M.; De Angelis, F.; de la Prida, L. M.; Valiente, M.; D'Orazio, A.; De Vittorio, M.; Grande, M.; Pisanello, F. Plasmonics on a Neural Implant: Engineering Light–Matter Interactions on the Nonplanar Surface of Tapered Optical Fibers. *Adv. Opt. Mater.* **2022**, *10* (2), 2101649. https://doi.org/10.1002/adom.202101649.

(50) Pisanello, F.; Sileo, L.; Oldenburg, I. A.; Pisanello, M.; Martiradonna, L.; Assad, J. A.; Sabatini, B. L.; De Vittorio, M. Multipoint-Emitting Optical Fibers for Spatially Addressable In Vivo Optogenetics. *Neuron* **2014**, *82* (6), 1245–1254. https://doi.org/10.1016/j.neuron.2014.04.041.

(51) D.G. Moore, H. R. T. Journal of Research of the National Bureau of Standards. *J. Res. Natl. Bur. Stand.* **1959**, *62*, 127–135.

(52) Mattox, D. M. Influence of Oxygen on the Adherence of Gold Films to Oxide Substrates. *J. Appl. Phys.* **1966**, *37* (9), 3613.

(53) Mohammad-Zadeh, L. F.; Moses, L.; Gwaltney-Brant, S. M. Serotonin: A Review. *Journal of Veterinary Pharmacology and Therapeutics* **2008**, *31* (3), 187–199. https://doi.org/10.1111/j.1365-2885.2008.00944.x.

(54) Zhao, S.; Piatkevich, K. D. Techniques for in Vivo Serotonin Detection in the Brain: State of the Art. *Journal of Neurochemistry* **2023**, *166* (3), 453–480. https://doi.org/10.1111/jnc.15865.

(55) Lu, R.; Konzelmann, A.; Xu, F.; Gong, Y.; Liu, J.; Liu, Q.; Xin, M.; Hui, R.; Wu, J. Z. High Sensitivity Surface Enhanced Raman Spectroscopy of R6G on in Situ Fabricated Au Nanoparticle/Graphene Plasmonic Substrates. *Carbon* **2015**.

(56) He, X. N.; Gao, Y.; Mahjouri-Samani, M.; Black, P. N.; Allen, J.; Mitchell, M.; Xiong, W.; Zhou, Y. S.; Jiang, L.; Lu, Y. F. Surface-Enhanced Raman Spectroscopy Using Gold-Coated Horizontally Aligned Carbon Nanotubes. *Nanotechnology* **2012**, *23* (20), 205702. https://doi.org/10.1088/0957-4484/23/20/205702.

(57) Bain, A. J.; Chandna, P.; Butcher, G.; Bryant, J. Picosecond Polarized Fluorescence Studies of Anisotropic Fluid Media. II. Experimental Studies of Molecular Order and Motion in Jet Aligned Rhodamine 6G and Resorufin Solutions. *The Journal of Chemical Physics* **2000**, *112* (23), 10435–10449. https://doi.org/10.1063/1.481679.

(58) López Arbeloa, F.; Martínez Martínez, V.; Arbeloa, T.; López Arbeloa, I. Photoresponse and Anisotropy of Rhodamine Dye Intercalated in Ordered Clay Layered Films. *Journal of Photochemistry and Photobiology C: Photochemistry Reviews* **2007**, *8* (2), 85–108. https://doi.org/10.1016/j.jphotochemrev.2007.03.003.

(59) Lyu, Y.; Becerril, L. M.; Vanzan, M.; Corni, S.; Cattelan, M.; Granozzi, G.; Frasconi, M.; Rajak, P.; Banerjee, P.; Ciancio, R.; Mancin, F.; Scrimin, P. The Interaction of Amines with Gold Nanoparticles. *Advanced Materials* *n/a* (n/a), 2211624. https://doi.org/10.1002/adma.202211624.

(60) Rocchigiani, L.; Bochmann, M. Recent Advances in Gold(III) Chemistry: Structure, Bonding, Reactivity, and Role in Homogeneous Catalysis. *Chem. Rev.* **2021**, *121* (14), 8364–8451. https://doi.org/10.1021/acs.chemrev.0c00552.

(61) Feng, J.; B. Pandey, R.; J. Berry, R.; L. Farmer, B.; R. Naik, R.; Heinz, H. Adsorption Mechanism of Single Amino Acid and Surfactant Molecules to Au {111} Surfaces in Aqueous Solution: Design Rules for Metal-Binding Molecules. *Soft Matter* **2011**, *7* (5), 2113–2120. https://doi.org/10.1039/C0SM01118E.

(62) Hoefling, M.; Iori, F.; Corni, S.; Gottschalk, K.-E. The Conformations of Amino Acids on a Gold(111) Surface. *ChemPhysChem* **2010**, *11* (8), 1763–1767. https://doi.org/10.1002/cphc.200900990.

(63) Zhong, Z.; Patskovskyy, S.; Bouvrette, P.; Luong, J. H. T.; Gedanken, A. The Surface Chemistry of Au Colloids and Their Interactions with Functional Amino Acids. *J. Phys. Chem. B* **2004**, *108* (13), 4046–4052. https://doi.org/10.1021/jp037056a.

(64) Lowe, A. B. Thiol-Ene "Click" Reactions and Recent Applications in Polymer and Materials Synthesis. *Polym. Chem.* **2010**, *1* (1), 17–36. https://doi.org/10.1039/B9PY00216B.

(65) Do, P. Q. T.; Huong, V. T.; Phuong, N. T. T.; Nguyen, T.-H.; Ta, H. K. T.; Ju, H.; Phan, T. B.; Phung, V.-D.; Trinh, K. T. L.; Tran, N. H. T. The Highly Sensitive Determination of Serotonin by Using Gold Nanoparticles (Au NPs) with a Localized Surface Plasmon Resonance (LSPR) Absorption Wavelength in the Visible Region. *RSC Adv.* **2020**, *10* (51), 30858–30869. https://doi.org/10.1039/D0RA05271J.

(66) Wang, P.; Xia, M.; Liang, O.; Sun, K.; Cipriano, A. F.; Schroeder, T.; Liu, H.; Xie, Y.-H. Label-Free SERS Selective Detection of Dopamine and Serotonin Using Graphene-Au Nanopyramid Heterostructure. *Anal. Chem.* **2015**, *87* (20), 10255–10261. https://doi.org/10.1021/acs.analchem.5b01560.

(67) Moody, A. S.; Sharma, B. Multi-Metal, Multi-Wavelength Surface-Enhanced Raman Spectroscopy Detection of Neurotransmitters. *ACS Chem. Neurosci.* **2018**, *9* (6), 1380–1387. https://doi.org/10.1021/acschemneuro.8b00020.

(68) Qiu, C.; Bennet, K. E.; Tomshine, J. R.; Hara, S.; Ciubuc, J. D.; Schmidt, U.; Durrer, W. G.; McIntosh, M. B.; Eastman, M.; Manciu, F. S. Ultrasensitive Detection of Neurotransmitters by Surface Enhanced Raman Spectroscopy for Biosensing Applications. **2017**, *7* (1), 1921–1926.

(69) Choi, H.-K.; Park, W.-H.; Park, C.-G.; Shin, H.-H.; Lee, K. S.; Kim, Z. H. Metal-Catalyzed Chemical Reaction of Single Molecules Directly Probed by Vibrational Spectroscopy. *J. Am. Chem. Soc.* **2016**, *138* (13), 4673–4684. https://doi.org/10.1021/jacs.6b01865.

(70) Shin, H.-H.; Jeong, J.; Nam, Y.; Lee, K. S.; Yeon, G. J.; Lee, H.; Lee, S. Y.; Park, S.; Park, H.; Lee, J. Y.; Kim, Z. H. Vibrationally Hot Reactants in a Plasmon-Assisted Chemical Reaction. *J. Am. Chem. Soc.* **2023**, *145* (22), 12264–12274. https://doi.org/10.1021/jacs.3c02681.

(71) Kazuma, E. Key Factors for Controlling Plasmon-Induced Chemical Reactions on Metal Surfaces. *J. Phys. Chem. Lett.* **2024**, *15* (1), 59–67. https://doi.org/10.1021/acs.jpclett.3c03120.

(72) Masson, J.-F.; Wallace, G. Q.; Asselin, J.; Ten, A.; Hojjat Jodaylami, M.; Faulds, K.; Graham, D.; Biggins, J. S.; Ringe, E. Optoplasmonic Effects in Highly Curved Surfaces for Catalysis, Photothermal Heating, and SERS. *ACS Appl. Mater. Interfaces* **2023**, *15* (39), 46181–46194. https://doi.org/10.1021/acsami.3c07880.

(73) Olmon, R. L.; Slovick, B.; Johnson, T. W.; Shelton, D.; Oh, S.-H.; Boreman, G. D.; Raschke, M. B. Optical Dielectric Function of Gold. *Phys. Rev. B* **2012**, *86* (23), 235147. https://doi.org/10.1103/PhysRevB.86.235147.

# Supporting Information for
# Tunable Nanoislands Decorated Tapered Optical Fibers Reveal Concurrent Contributions in Through-Fiber SERS Detection

*Di Zheng[1,\*], Muhammad Fayyaz Kashif[1], Linda Piscopo[1,2], Liam Collard[1,3], Cristian Ciracì[1], Massimo De Vittorio[1,2,3,\*,†], Ferruccio Pisanello[1,3,\*,†]*

[1] Istituto Italiano di Tecnologia, Center for Biomolecular Nanotechnologies, Arnesano, LE 73010, Italy
[2] Dipartimento di Ingegneria Dell'Innovazione, Università del Salento, Lecce 73100, Italy
[3] RAISE Ecosystem, Genova, Italy

*Authors to whom correspondence should be addressed: di.zheng@iit.it, massimo.devittorio@iit.it, Ferruccio.pisanello@iit.it
† These authors contributed equally to this work

- **S1 Tunable nanoislands (NIs) morphology analysis**

As the initial Au film thickness increases, the diameter and inter-NIs distances of fabricated patterns also increase. To quantitatively analyze the SEM images of different patterns, we developed our custom software, the user interface is shown in Fig.S1b. The software allowed us to load the SEM image (Fig.S1a is an example of NIs patterns fabricated with 5 nm Au film deposition), automatically read the scale bar for each image, and display the gray-level histogram. we first examine the gray-level histogram of the imported image, the distribution is usually two peaks (Fig.S1b). By choosing the proper gray-level threshold between the two peaks, we transfer the SEM image into binary images with the shape of the NIs occupation. For the patterns fabricated with initial Au film thickness of 15 nm and 20 nm, the shape of the NIs is less round, and the featured dimensions of the structures can be well resolved in the SEM images, so the particle count and the coverage rate can be obtained by tracing object boundaries. The parameter of the minimal area was set to exclude the boundaries that are too small to be NIs. For the NIs patterns fabricated with 5 nm and 10 nm Au film coating, the NIs are mostly round, and the featured dimensions of the nanostructure are small, especially since the inter-NIs distances are approaching the limit of SEM resolution. Thus, the particle count obtained by tracing object boundaries (Fig.S1d) can underestimate the actual number of the NIs under analysis, as the method cannot reliably distinguish the NIs located very close to each other (gap <10 nm). For the reliable count of the particle numbers, we use the circular Hough Transform to recognize the subround objects, the circle radius range (MinRdii, MaxRadii) was finely adjusted to optimize the accuracy of the particle count, as shown in Fig.S1c. For all types of patterns, the coverage rate was analyzed by tracing object boundaries (Fig.S1d), and then using the summation of the areas of the boundaries divided by the total surface that had been analyzed. For each fiber, the SEM images included for analysis are no less than 10 and have different magnifications. The extracted geometric parameters, such as average diameter and the coverage rate were the mean results of all the images that have been analyzed.

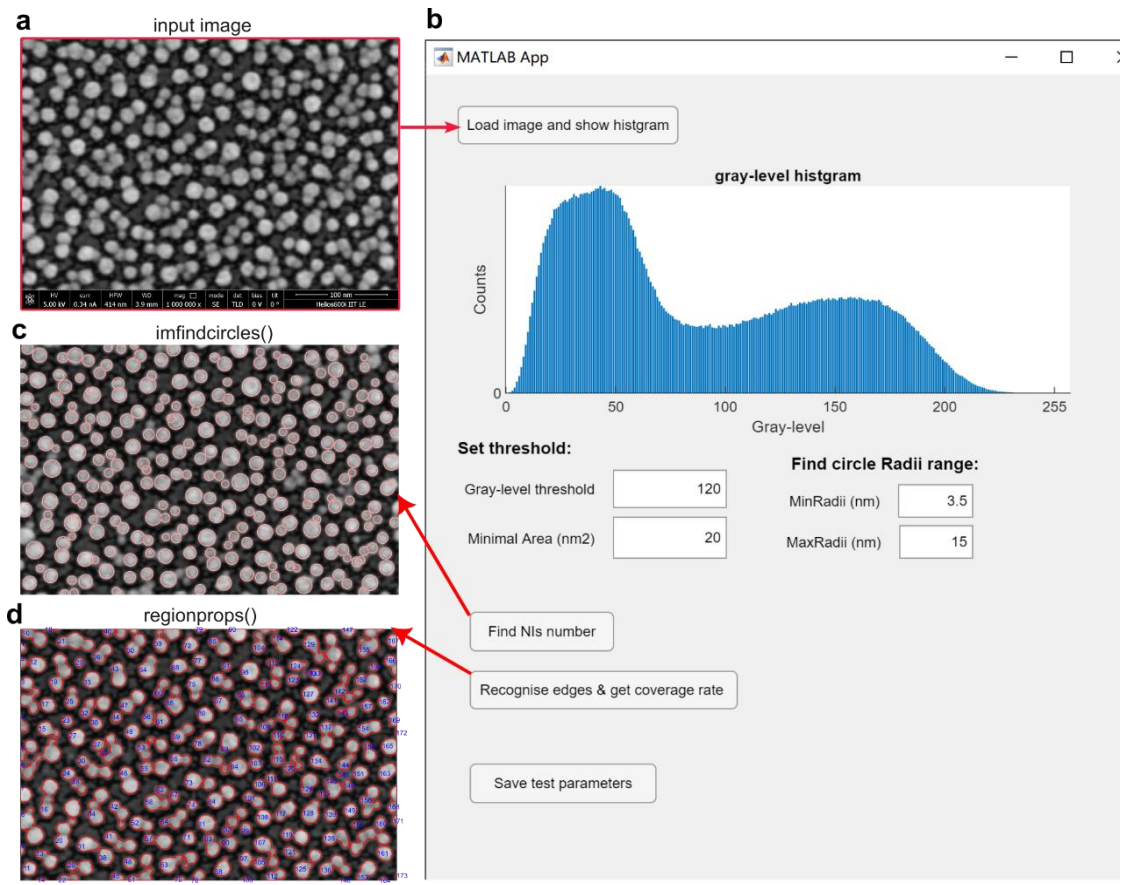

***Figure S1*** *The morphological analysis of the tunable NIs-TFs. (a) Representative morphology of NIs fabricated with 5 nm Au deposition. (b) The user interface of the custom software that has been used to analyze NIs patterns. (c) The NIs analysis with circular Hough transforms methods. (d) The NIs analysis with tracing object boundaries methods.*

- **S2 The electromagnetic simulations**

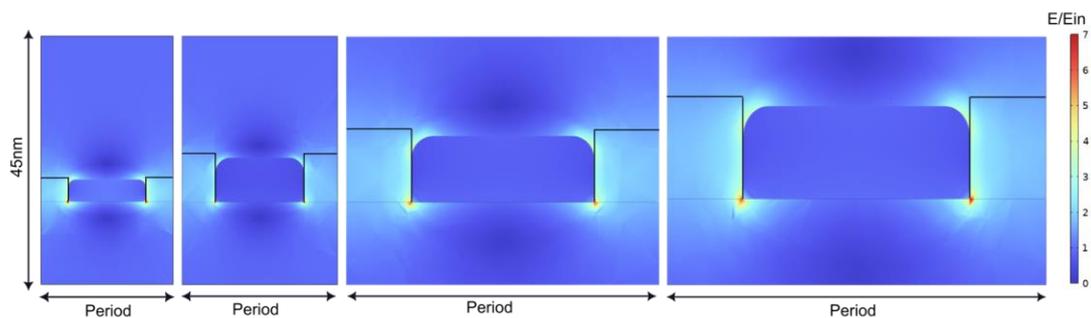

***Figure S2*** *The cross-sections of field-enhancement distributions for the hexagonal periodic arrays (*D14H4G7, D16H8G8, D33H12G16, and D49H16G22*). The field enhancement is mainly distributed at the edges and within the gaps, the black lines outline the regions integrated into volume field enhancement for each unit cell pattern.*

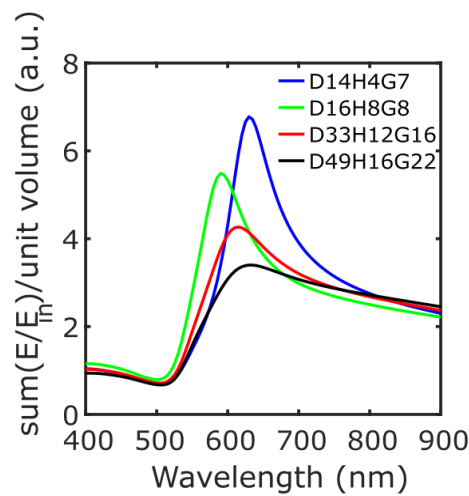

*Figure S3* The summation of near-field enhancement normalizes to integrated volume along with different wavelengths produced by each morphology as a function of the incident wavelength under normal incidence.

- **S3 The optical setup for the limit of detection measurements**

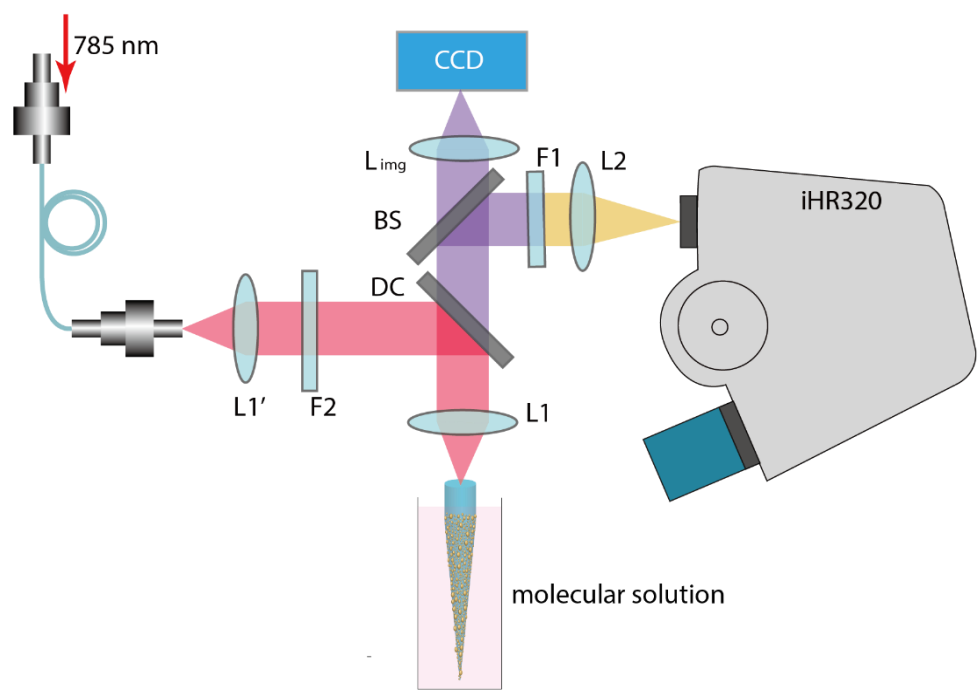

*Figure S4* The optical setup for Raman measurements with different sample configurations at 785 nm. DC (dichroic mirror, Semrock, LPD02-785RU-25), focus lens L1 and L1' (aspheric Ø25.0 mm, f = 20 mm), collection lens L2 (Thorlabs, TTL200MP, f = 200 mm), image lens $L_{img}$ (Thorlabs, WFA4100), F1 (long pass filter, Semrock, LP02-785RU-25), F2 (laser line filter, Semrock, LL01-785-25), BS (beam splitter with T : R = 50 : 50), CCD (Thorlabs CMOS camera, DCC1545M).

- **S4 Spectra data analysis for R6G and serotonin**

In Fig.S4, we show a representative data analysis on the R6G limit of detection (LOD) SERS spectra set measured from a D49H16G22 NIs-TF. All the other spectra sets were treated the same way. In the manuscript, we show the original spectra and the substrate spectra, to obtain the bubble chart, and make the peak area integral intensity comparable across different fibers with different patterns, the normalization procedure was used. We first normalized the original spectra set to the silica peak at 1055 $cm^{-1}$ (Fig.S4c), and then we subtracted the normalized silica background (spectra measured in water) within the normalized spectra, obtaining the subtracted normalized spectra in Fig.S4d. Eventually, the peak area integral was calculated from Fig.S4d. For each peak, a local linear baseline was used to remove the background. For each set of spectra data measured from one fiber, we can obtain a set of peak area integral data shown in Fig.S4e. Each bubble chart in the manuscript is the union results from two data sets obtained from two fibers with the same NIs patterns. The intensity thresholds of 0.19 for R6G data and 0.06 for serotonin data were first used to roughly eliminate the analyzed positions without peak, then a logic mask was applied to the matrix data for accurate elimination of the positions without peak. The logic masks were obtained by carefully examining the subtracted spectra and original spectra manually across the data sets.

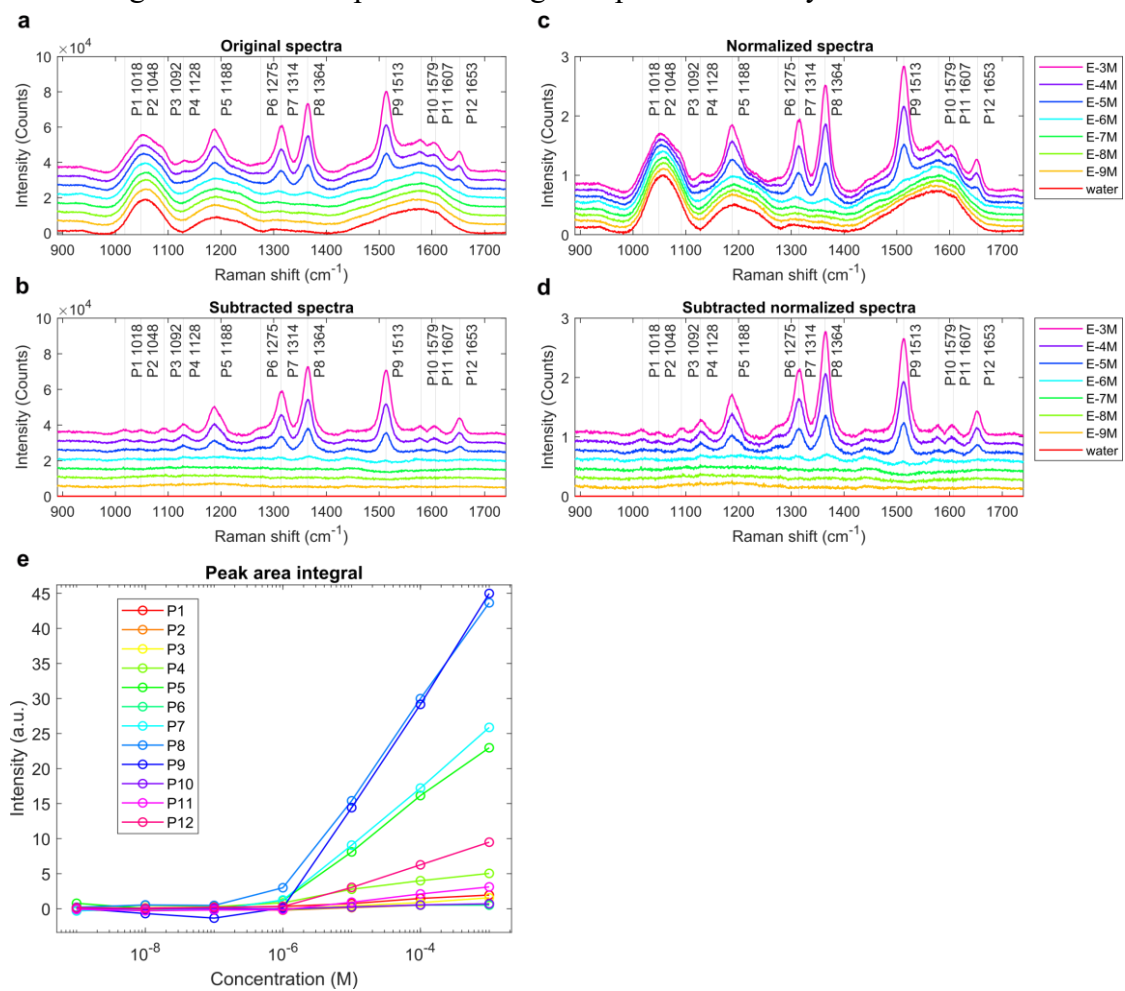

*Figure S5* *Spectra analysis of an R6G limit of detection SERS spectra set measured from a D49H16G22 NIs-TF. (a) Displays the original spectra, ranging from 0 to $10^{-3}$*

*M. (b) Subtracted spectra obtained by eliminating the silica background measured in water. (c) The normalized spectra were obtained by normalizing the original spectra to the silica peak at 1055 cm$^{-1}$. (d) Subtracted normalized spectra, further refined by subtracting the silica background in (c), all spectra have been vertically offset for clarity (e) Integrated peak areas were calculated from the spectra in (d), and plotted against concentrations.*

In the manuscript, we demonstrate that the NIs-TFs measured with serotonin exhibit degraded sensitivity as the concentration increases. Following this degradation, the same fiber depicted in Fig. 5d was utilized to conduct the LOD experiment for R6G. The results, presented in Fig. S5, indicate that the fiber is still capable of detecting R6G molecules, albeit with a reduced peak intensity compared to the fresh sample (refer to Fig. S4e), and with a one-order decrease in detection sensitivity.

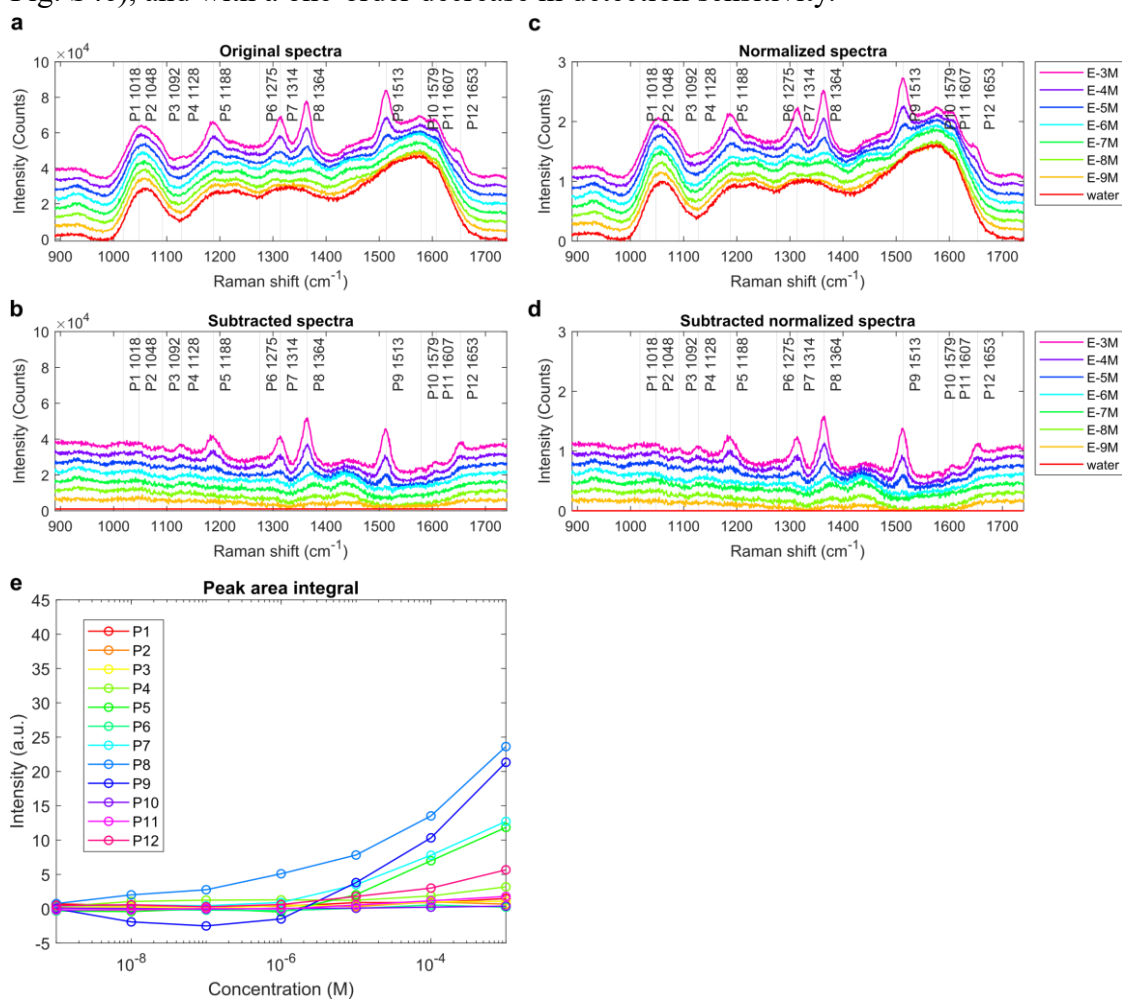

*Figure S6 Spectra analysis of an R6G limit of detection SERS spectra set measured with D49H16G22 NIs-TF after serotonin measurement. (a) Displays the original spectra, ranging from 0 to 10$^{-3}$ M. (b) Subtracted spectra obtained by eliminating the silica background measured in water. (c) The normalized spectra were obtained by normalizing the original spectra to the silica peak at 1055 cm$^{-1}$. (d) Subtracted normalized spectra, further refined by subtracting the silica background in (c), all spectra have been vertically offset for clarity (e) Integrated peak areas were calculated from the spectra*

*in (d) and plotted against concentrations.*

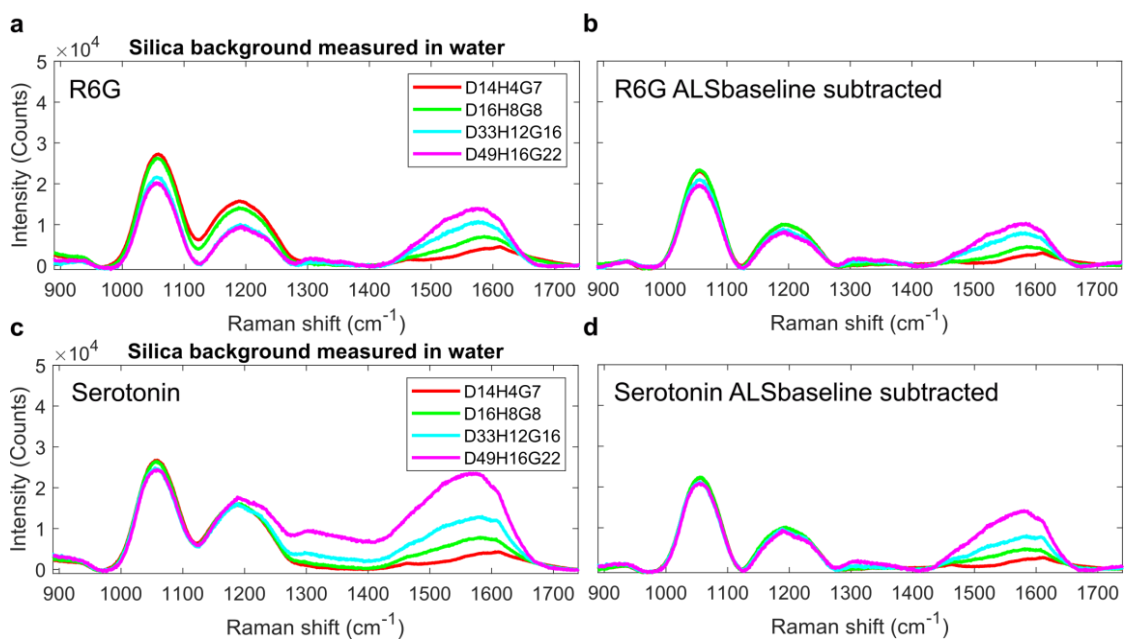

*Figure S7 The silica backgrounds for tunable NIs-TF set. (a) The silica backgrounds for fibers in R6G LOD measurements in the manuscript and (b) the corresponding ALS baseline subtracted spectra. Each spectrum is an average between two backgrounds from two fibers with the same pattern. (c) The silica backgrounds for fibers in serotonin LOD measurements in the manuscript and (d) the corresponding ALS baseline subtracted spectra. Each spectrum is an average between two backgrounds from two fibers with the same pattern.*